\documentclass{ametsocV6.1}

\usepackage{anyfontsize}
\renewcommand{\vec}[1]{\mathbf{#1}}

\usepackage{acro}
\DeclareAcronym{sst}{
  short=SST,
  long=sea surface temperature,
}
\DeclareAcronym{slp}{
  short=SLP,
  long=sea level pressure,
}
\DeclareAcronym{amoc}{
  short=AMOC,
  long=Atlantic Meridional Overturning Circulation,
}
\DeclareAcronym{nao}{
  short=NAO,
  long=North Atlantic Oscillation,
  long-format=\itshape,
}
\DeclareAcronym{enso}{
  short=ENSO,
  long=El Niño Southern Oscillation,
  long-format=\itshape,
}
\DeclareAcronym{mjo}{
  short=MJO,
  long=Madden-Julian Oscillation,
  long-format=\itshape,
}
\DeclareAcronym{qbo}{
  short=QBO,
  long=Quasi-Biennial Oscillation,
  long-format=\itshape,
}
\DeclareAcronym{itcz}{
  short=ITCZ,
  long=Intertropical Convergence Zone,
  long-format=\itshape
}
\DeclareAcronym{gcm}{
  short=GCM,
  long=General Circulation Model,
  long-format=\itshape,
}
\DeclareAcronym{cesm}{
  short=CESM,
  long=Community Earth System Model,
  long-format=\itshape,
}

\title{Factors governing the existence of an abrupt transition to superrotation in an idealized GCM}

\authors{Corentin Herbert\aff{a}\correspondingauthor{Corentin Herbert, corentin.herbert@ens-lyon.fr}}

\affiliation{\aff{a}{CNRS, ENS de Lyon, LPENSL, UMR 5672, Lyon, France}}

\abstract{Some numerical simulations of very warm climates suggest that the Earth's atmosphere may undergo a transition to a state of equatorial superrotation, where the zonal-mean zonal wind in the tropics is westerly. However, major uncertainties remain about the circumstances under which such a transition could happen. A natural first step towards reducing these uncertainties is to better understand the dynamical processes involved in the transition in idealized setups. However, simple numerical experiments have reported very different responses to tropical diabatic heating in different models, with both a continuous and an abrupt transition to superotation. In this paper, we investigate the mechanisms controlling the nature of the transition. We show that in an idealized Held-Suarez framework, it is governed by both the meridional temperature gradient and the bottom friction coefficient. These two parameters control a competition between two feedback mechanisms: a positive tropical wave-jet mechanism, and a negative feedback mechanism related to absorption of extratropical waves near their critical latitudes in the tropics.}

\begin{document}

\maketitle

%
%
%
%
%
%

%








\section{Introduction}


In the current climate, climatological winds in the tropical troposphere are easterly~\citep{Lee1999}, unlike many other planetary atmospheres, such as Jupiter, Saturn, Venus or Titan, which exhibit a strong westerly jet at the equator~\citep{Read2018,Imamura2020}.
The latter case, referred to as superrotation, occurs seasonally on Earth~\citep{Zhang2022}.
It has been suggested that under different circumstances, in particular in warm climates~\citep{Pierrehumbert2000,Tziperman2009}, the Earth might be in permanent superrotation.
For instance, \citet{Caballero2010,Lan2023,Zurita-Gotor2025} observed in numerical simulations with \acp{gcm} a spontaneous transition to superrotation for high atmospheric CO\textsubscript{2} concentrations.
The reversal of the direction of the mean wind can be expected to affect many aspects of tropical dynamics.
\citet{Tziperman2009} suggested early on an interplay with El Niño, but the numerical experiments of~\citet{Caballero2018} revealed that a direct coupling with the ocean through a reversal of the zonal winds at the surface was only obtained with strongly enhanced convective momentum transport.
On the other hand, a connection with the properties of the \ac{mjo} has been emphasized by~\citet{Arnold2013a,Carlson2016}.

However, a broader impact on the mean climate could also occur through feedback processes, potentially leading to state-dependent climate sensitivity~\citep{Caballero2013}.
Recently, \citet{Marino2026} showed in idealized \ac{gcm} simulations that even without any radiative forcing, a transition to superrotation would result in a significant global warming at the surface (comparable to CO\textsubscript{2} doubling in these experiments), exacerbated in the mid-latitudes, and a major reorganization of the tropical water cycle, with a collapse of the \ac{itcz} and a reduced meridional gradient of relative humidity.
These findings suggest first that superrotation might act as a feedback mechanism on top of a given warming due to radiative forcing, but also that the impacts of a warming reaching the threshold of the transition to superrotation would be strongly amplified by the response of the large-scale tropical circulation.
For these reasons, it appears desirable to improve our understanding of the conditions under which superrotation could set in on an Earth-like planet, both due to its potential importance for interpreting paleoclimate proxies for warm climates, and to better assess whether it might be relevant for projections of future climates.
Because current models simulate a spontaneous transition to superrotation only for very strong forcing (in regimes where many parameterizations might no longer be accurate), it is difficult to estimate to what extent these results are reliable.
One way to make progress in this direction is through a better theoretical understanding of the mechanisms governing the transition to superrotation.

Previous studies have already uncovered multiple routes to superrotation.
All of them require some anti-diffusive transport of angular momentum towards the equator, which can only be achieved by eddy fluxes.
However, the mechanisms generating these fluxes may differ.
A number of studies considered a small-scale vorticity source directly in the tropics, often in the context of the shallow-water equations~\citep{Scott2008,Saito2015,Warneford2017,Suhas2017,Schrottle2022}.
Alternatively, motivated by planetary atmospheres, \citet{Iga2005,Mitchell2010} on the one hand and~\citet{Williams2003,Williams2006} on the other hand have shown that the eddies transporting momentum towards the equator, resulting in superrotation, could also be due to hydrodynamical instabilities, such as the Kelvin-Rossby instability~\citep{Wang2014b,Potter2014,Zurita-Gotor2018a,Zurita-Gotor2022, Zurita-Gotor2024}.
Finally, a possibility which was suggested early on by~\citet{Suarez1992} and~\citet{Saravanan1993} and followed-up by~\citet{Kraucunas2005,Showman2010,Arnold2012,Laraia2015} and~\citet{Lutsko2018} is that the eddies might be generated as a response to zonal anomalies of diabatic heating.

In spite of our knowledge of these dynamical pathways in idealized settings, some questions still hinder our ability to reliably estimate the possibility that superrotation appears spontaneously in warm climates.
For instance, the importance of the base state, and the model-dependence of the response to idealized forcing remains unclear.
In addition, it has been suggested that, at least in some of the idealized experiments cited above, the response to forcing might exhibit some memory effects, apparent for instance through hysteresis phenomena.
This is an important problem for multiple reasons: first, if the transition between a sub-rotating and a superrotating circulation is indeed governed by a bifurcation, this provides a mechanism for abrupt climate change, either when the forcing crosses the stability threshold, or when internal fluctuations generate noise-driven transitions between the metastable states.
Second, it matters for the interpretation of \ac{gcm} simulations: if the transition is smooth, it is easier to detect that increasing levels of forcing have an increasing effect on the tropical winds.
On the other hand, for an abrupt transition the response is often small until approaching the bifurcation.

In the early simulations of~\citet{Suarez1992} and~\citet{Saravanan1993} using two-level models with large-scale diabatic heating, the transition to superrotation was indeed found to be abrupt.
In a multi-level model with similar forcing, \citet{Arnold2012} also found an abrupt transition, and suggested that it was governed by a wave-jet resonance mechanism.
\Citet{Herbert2020} studied in details this mechanism as well as the Hadley cell feedback introduced by~\citet{Shell2004}, to show that both lead to multiple equilibria in low-dimensional models, but only the former appears to be robust in a multi-level model.
Nevertheless, it remains unclear under which conditions exactly a bifurcation exists, and to what extent it is robust across different dynamical cores.
Indeed, using a different model, \citet{Lutsko2018} did not report any sign of bifurcation in a very similar setup over a broad range of imposed tropical diabatic heating amplitudes.

The goal of this paper is therefore to investigate, within the context of a single model, the factors controlling the nature of the transition to superrotation and the robustness of the existence of a bifurcation leading to hysteresis behavior.
We do so in an idealized framework which allows us to identify a restricted number of key parameters: the meridional temperature gradient at radiative equilibrium and the bottom friction.
In section~\ref{sec:model}, we present this framework and the design of our numerical experiments, consisting mostly of hysteresis experiments.
The result of these experiments are first described in section~\ref{sec:parameters}, emphasizing the role of the control parameters we have identified.
In particular, we find that, with the same model as~\citet{Lutsko2018}, both abrupt and continuous transitions to superrotation are possible depending on these parameters.
Then in section~\ref{sec:mechanisms}, we proceed with a more detailed analysis of the feedback mechanisms governed by these controlled parameters, and how they compete to determine whether the transition to superrotation is smooth or abrupt.
Finally, our conclusions are presented in section~\ref{sec:conc}.

\section{Methodology} \label{sec:model}

\subsection{Numerical model: dynamical core}

Isca~\citep{Vallis2018} is an idealized atmospheric GCM based on the GFDL dynamical core~\citep{Gordon1982}.
It solves the primitive equations in vorticity-divergence form using sigma coordinates on the vertical~\citep{Bourke1974}.
The discretization is done using a triangular spectral truncature on the horizontal, and a centered difference scheme on the vertical.
Although the model implements the hybrid pressure-sigma coordinates of~\cite{Simmons1981}, we shall only use the pure sigma coordinates version.
The dynamical core is the same as the one in the FMS model used for instance in~\citet{Schneider2006a} or~\citet{Potter2014}.
In this paper, we use the dry version of Isca.

Hence, the model equations are the following:
\begin{align}
  \frac{D\vec{u}}{Dt} + f\vec{k} \times \vec{u}  &= - \nabla_\sigma \Phi - R_d T\nabla q + \vec{F}, \\
  \partial_\sigma \Phi &= - \frac{R_d T}{\sigma},\\
  \partial_t q &= - \tilde{\vec{u}}\cdot \nabla q - \nabla \cdot \tilde{\vec{u}},\\
  \frac{DT}{Dt} &= \frac{R_d T}{c_p p} \omega + Q,\\
    \dot{\sigma} &= -\int_0^\sigma (\vec{u}-\tilde{\vec{u}}) \cdot \nabla q - \int_0^\sigma (\nabla_\sigma \cdot \vec{u} - \nabla \cdot \tilde{\vec{u}}),\\
    \omega &= p \left\lbrack \frac{\dot{\sigma}}{\sigma} + (\vec{u}-\tilde{\vec{u}})\cdot \nabla q - \nabla \cdot \tilde{\vec{u}}\right\rbrack,
\end{align}
where $\vec{u}$ is the horizontal velocity, $\dot{\sigma}=\frac{D\sigma}{Dt}$ the vertical velocity in sigma coordinates, $\omega=\frac{Dp}{Dt}$ the vertical velocity in pressure coordinates, $\Phi$ the geopotential, $T$ the temperature, $p$ the pressure, and $q$ the logarithm of surface pressure.
$\tilde{\vec{u}}=\int_0^1\vec{u}d\sigma$ is the vertically-averaged (mass-weighted) horizontal velocity.
$Q$ represents diabatic heating and $F$ dissipative terms, described below.

All the runs shown here use a T42 resolution, with 25 vertical levels.

\subsection{Numerical model: physical parameterizations}\label{sec:hs}

The physics is represented in an idealized way, following~\cite{Held1994} and subsequent idealized primitive equations GCM studies.

The divergence of the wind stress tensor, $\vec{F}$, only contains a horizontal hyperdiffusion term $\vec{F}_\nu$, and Rayleigh damping for low-level winds: $\vec{F}_r = - \vec{u}/\tau_F$, with $\tau_F^{-1}=\tau_f^{-1}\max\left(0, \frac{\sigma-\sigma_b}{1-\sigma_b}\right)$, $\tau_f=1$ day and $\sigma_b=0.7$.
An additional damping term may be included to dissipate energy in a \emph{sponge layer} at the upper boundary; we do not use such a dissipation term here.

The model is forced by prescribing the diabatic heating $Q$.
It is the sum of two terms: $Q=Q_{HS}+Q_{trop}$.
The first term is the standard Newtonian relaxation towards a prescribed radiative equilibrium temperature field~\citep{Held1994}:
\begin{equation}
Q_{HS} = - \frac{T-T_{eq}}{\tau_Q}, \quad T_{eq}(p,\phi) = \max\left(200 K, \left\lbrack 315 K - \Delta_h \sin^2\phi - \Delta_v \ln(p/p_0) \cos^2\phi \right\rbrack {(p/p_0)}^{R/c_p} \right).
\end{equation}
This heating term is axisymmetric.
We use $\Delta_h=40$ K, $\Delta_v = 10$ K and $p_0 = 1000$ hPa as reference values.
The relaxation timescale depends on latitude and vertical level: $\tau_Q^{-1} = \tau_a^{-1}+(\tau_s^{-1}-\tau_a^{-1})\max\left(0, \frac{\sigma-\sigma_b}{1-\sigma_b}\right)\cos^4\phi$, and $\tau_a=$ 40 days, $\tau_s=4$ days.
The second term is a non-zonal heating restricted to the tropics (with a Gaussian profile), with a broad vertical distribution peaking in the mid-troposphere:
\begin{equation}
Q_{trop}(x, \phi, p, t) = Q_0\cos(k_f(x-c_f t))e^{-{(\phi/\Delta \phi)}^2}\max\left(\sin\left(\frac{p-p_t}{p_b-p_t}\pi\right), 0\right).
\end{equation}
All the runs shown here use $\Delta \phi=10^\circ$, $p_t=200$ hPa and $p_b=800$ hPa.
A forcing of this form has been used in several superrotation studies~\citep{Suarez1992, Saravanan1993, Kraucunas2005, Arnold2012}.
\cite{Lutsko2018} considered a variant where a non-periodic localized heating is applied.
To the best of our knowledge, only~\cite{Arnold2012} have considered the propagating version ($c_f \neq 0$); in the present work we found that it did not affect the main conclusions so we only present simulations with a steady diabatic heating.

This version of the model uses no convection parameterization.

\subsection{Numerical experiments}

\begin{figure*}[tbhp]
  \centering
  \includegraphics[width=0.45\linewidth]{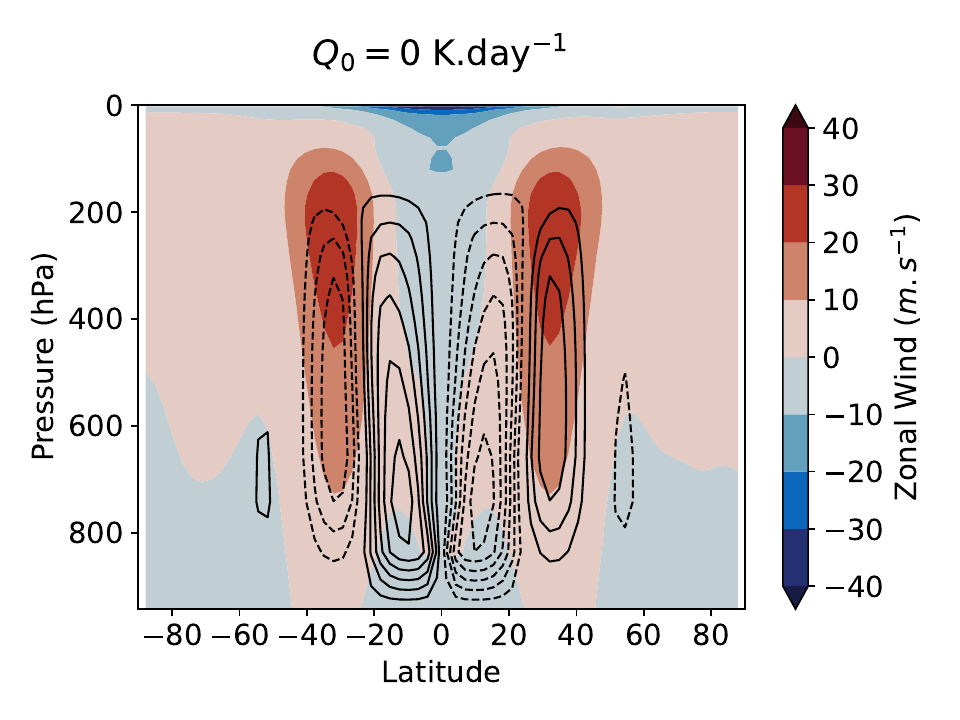}
  \includegraphics[width=0.45\linewidth]{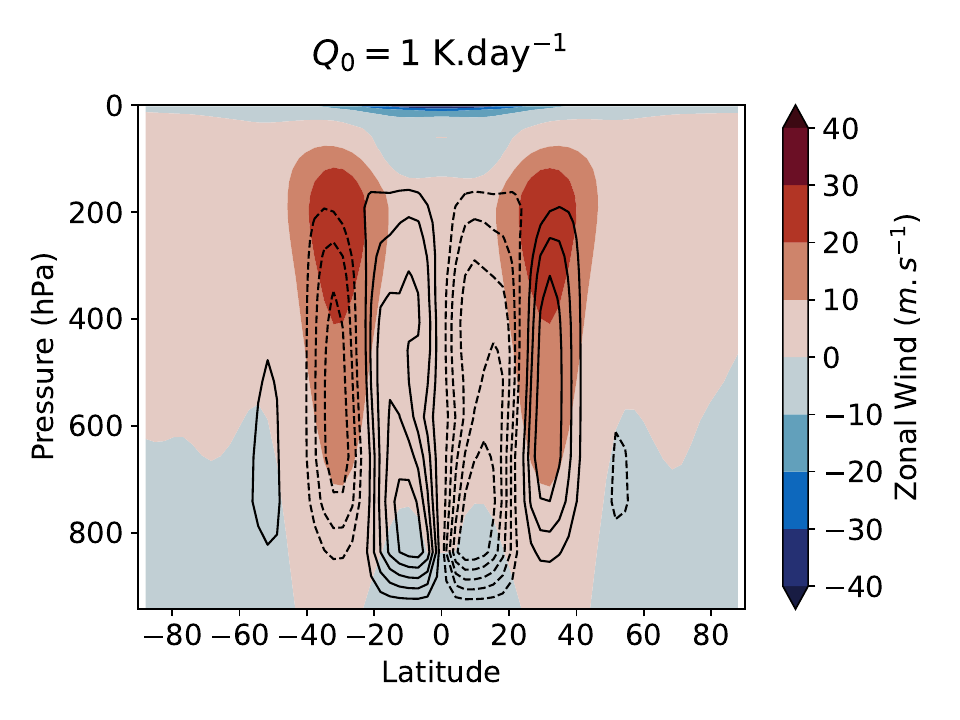}
  \includegraphics[width=0.45\linewidth]{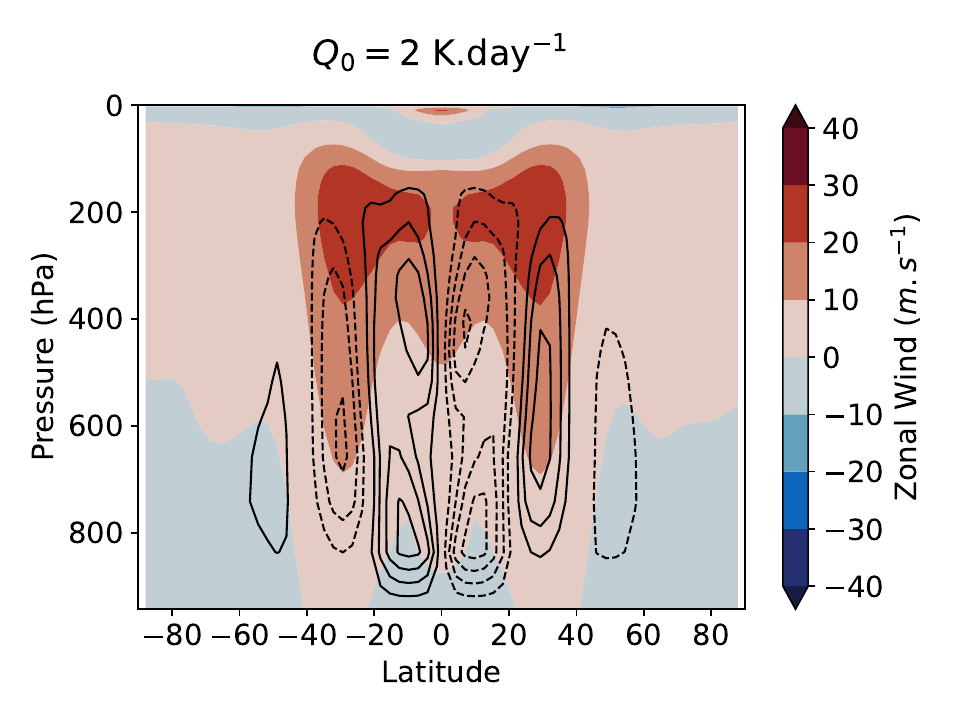}
  \includegraphics[width=0.45\linewidth]{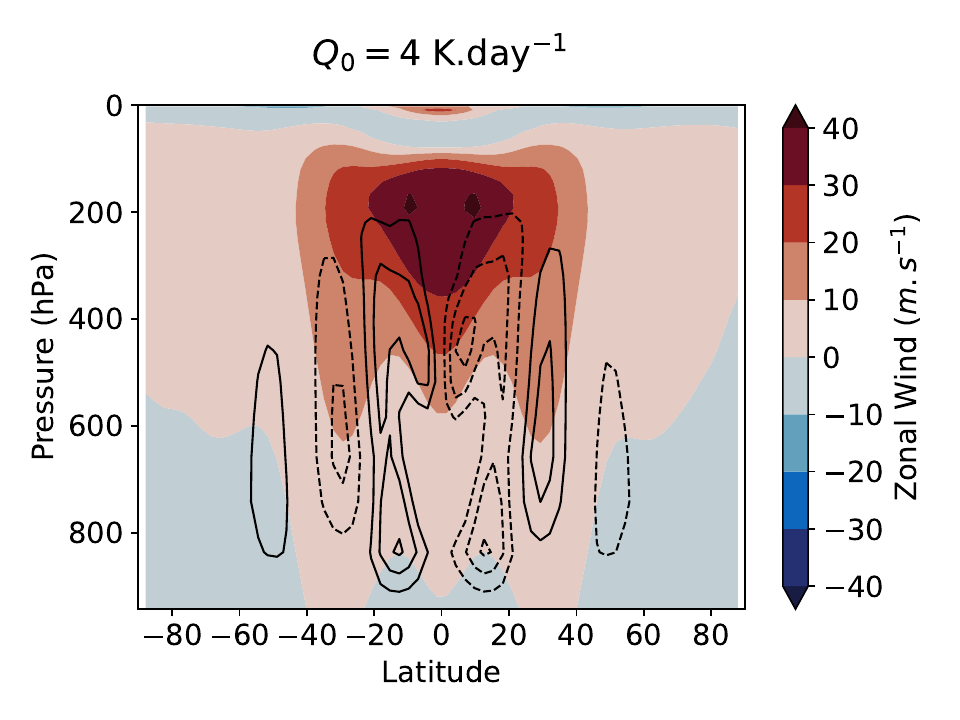}

  \caption{Zonally-averaged zonal wind (shading) and meridional mass streamfunction (contours) for different values of the tropical diabatic heating amplitude, from $Q_0=0$ (control run) to $Q_0=4$ K.day\textsuperscript{-1}. The contour interval for the meridional mass streamfunction is 10\textsuperscript{10} kg.s\textsuperscript{-1}. For all the runs in this figure the forcing parameters are $k_f=2$ and $c_f=0$.}
  \label{fig:circulation}
\end{figure*}

We carry out numerical simulations with increasing values of the tropical diabatic heating amplitude $Q_0$, from 0 to 4 K.day\textsuperscript{-1}.
The zonally-averaged circulation (zonal wind and meridional mass streamfunction) for four values of $Q_0$ are shown in Fig.~\ref{fig:circulation} (here we use the forcing wave number $k_f=2$ but the results presented in this section do not depend on this choice).
For $Q_0=0$ (control run, top-left panel of Fig.~\ref{fig:circulation}), we recover the standard features of Held-Suarez runs, which reproduce qualitatively the atmospheric circulation of the Earth, with in particular westerlies over most of the troposphere outside of the tropics, and easterlies over the whole depth of the atmosphere in the tropics, and near the surface at high latitudes.
We observe that increasing the amplitude of the non-zonal diabatic heating in the tropics leads to a state of equatorial superrotation, as previous studies with similar setups have already reported~\citep{Suarez1992, Saravanan1993, Kraucunas2005, Arnold2012, Lutsko2018}.
For intermediate values of the forcing amplitude, e.g. $Q_0=1$ K.day\textsuperscript{-1} (top right panel of Fig.~\ref{fig:circulation}), weak westerlies (below 10 m.s\textsuperscript{-1}) appear over most of the troposphere at the equator, corresponding to a weak superrotation.
Increasing further the forcing amplitude, e.g. $Q_0=2$ K.day\textsuperscript{-1} (bottom left panel of Fig.~\ref{fig:circulation}) leads to a stronger superrotation state, where the wind speed of the westerlies at the equator is comparable to the wind speed in the extratropical jets.
Finally, for still stronger tropical diabatic heating ($Q_0=4$ K.day\textsuperscript{-1}, bottom right panel of Fig.~\ref{fig:circulation}), the maximum wind speed is reached in the equatorial jet.
As the tropical diabatic heating amplitude increases, the strength of the Hadley cell progressively decreases, from a maximum value of the meridional mass streamfunction of 6.7$\times 10^{10}$ kg.s\textsuperscript{-1} in the control run to 3.4$\times 10^{10}$ kg.s\textsuperscript{-1} for $Q_0=4$ K.day\textsuperscript{-1}. Such a weakening of the meridional overturning circulation is comparable to the one observed in \ac{gcm} simulations of spontaneous transitions to superrotation~\citep{Zurita-Gotor2025}, and slightly weaker than the one obtained when superrotation is forced by imposing prescribed eddy fluxes~\citep{Marino2026}.

In the sequel, we carry out hysteresis experiments to investigate the nature of the transition as the tropical heating amplitude increases.
The goal is to understand whether the transition is continuous and reversible, or whether it is abrupt and can exhibit hysteresis, corresponding to the presence of a bifurcation.
We find that both cases are possible, and we investigate which parameters control the existence of bifurcations.
The results of these experiments are presented in Sec.~\ref{sec:parameters}, and we analyse the competing feedback mechanisms in more details in Sec.~\ref{sec:mechanisms}.

\section{Impact of the meridional temperature gradient and bottom friction on the transition to superrotation}\label{sec:parameters}

\subsection{Impact of the meridional temperature gradient on the transition to superrotation}\label{sec:meridional_gradient}

\begin{figure*}[tbhp]
  \centering
  \includegraphics[width=0.48\linewidth]{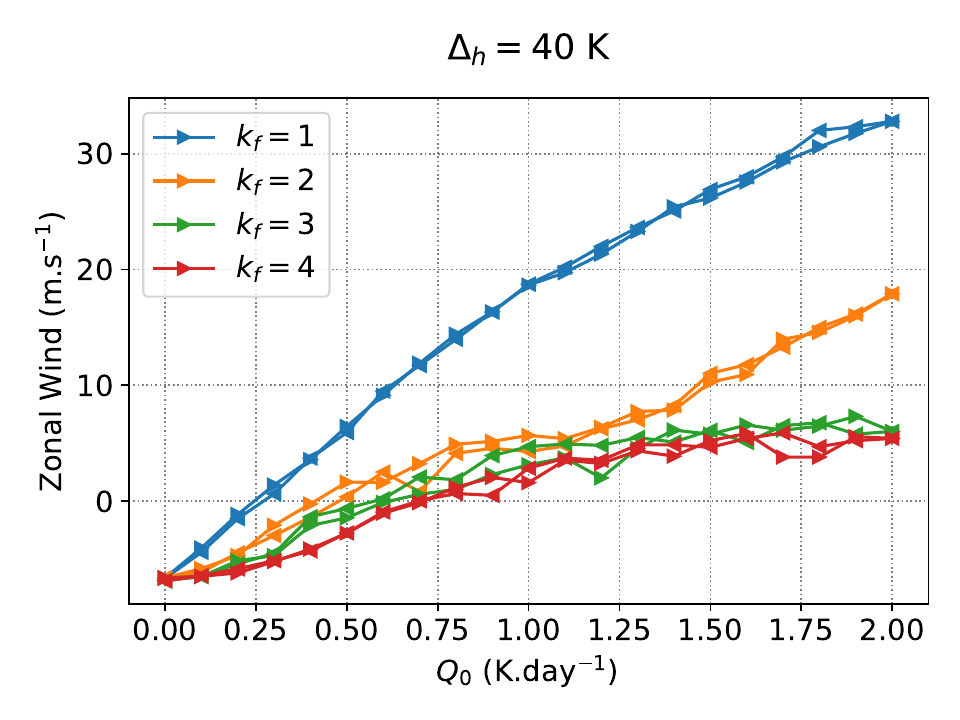}
  \includegraphics[width=0.48\linewidth]{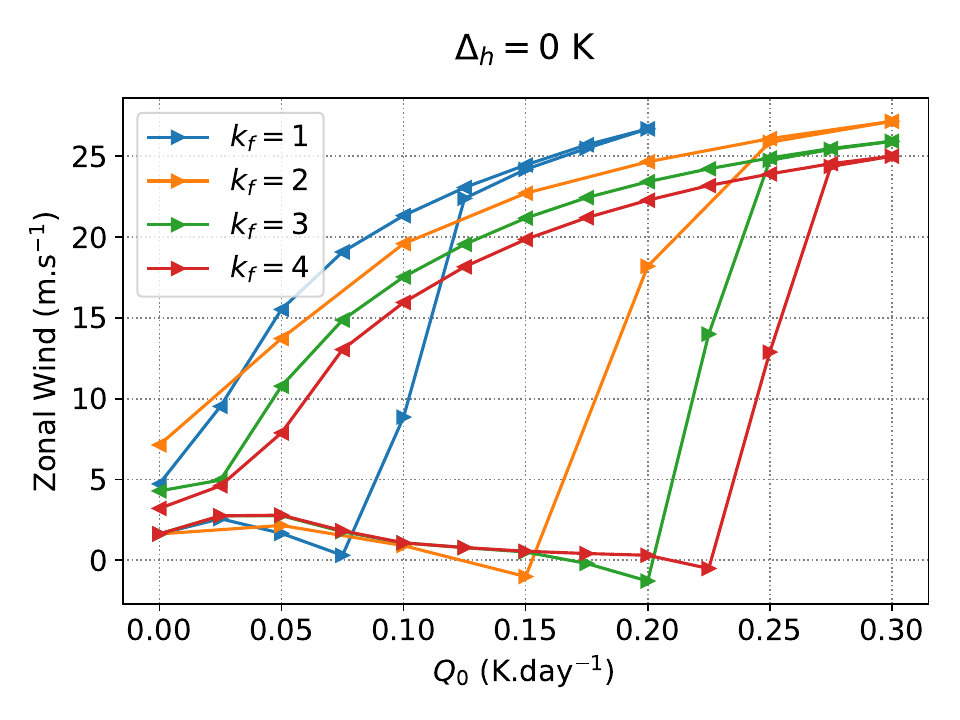}
  \caption{Zonally-averaged zonal wind in the upper troposphere at the equator as a function of the diabatic heating amplitude $Q_0$, for four different forcing wavenumbers ($k_f=1$, 2, 3 and 4) and two values of the meridional temperature gradient: $\Delta_h=40$ K (left) and $\Delta_h=0$ (right). Right-facing triangles represent points on the increasing $Q_0$ branch and left-facing triangles points on the decreasing $Q_0$ branch.}
  \label{fig:hysteresis_dh}
\end{figure*}

We first carry out hysteresis experiments, ramping up and down the amplitude of the non-zonal tropical heating $Q_0$, with other parameters fixed at their default values given above.
The results of these experiments are shown in Fig.~\ref{fig:hysteresis_dh} (left), for four different forcing wave numbers, $k_f=1,2,3$ and 4.
We proceed in a discontinuous manner, increasing $Q_0$ with finite increments $\Delta{}Q_0=0.1$ K.day\textsuperscript{-1} with a time step $\Delta t=5000$ days until we reach $Q_0=2$ K.day\textsuperscript{-1}.
After this value, we proceed similarly with opposite increments until going back to $Q_0=0$.
The value of the zonal wind is averaged over time and over the upper tropical troposphere (i.e. over the region $|\phi| \leq 5$\textdegree, 100 hPa $\leq p \leq$ 300 hPa).
This procedure allows us to check that the system relaxes to a new equilibrium at each step, so that the curves describe quasi-equilibrium states of the system, and dynamical hysteresis phenomena~\citep[e.g.]{Jung1990} are excluded.
Apart from small discrepancies, due to statistical fluctuations, the increasing $Q_0$ and the decreasing $Q_0$ branches collapse almost perfectly, indicating that the transition to superrotation is reversible in these experiments.
In particular, there is no hysteresis (the same conclusion applies for larger forcing wave numbers $k_f$ or for a propagating forcing with $c_f=5$ m.s\textsuperscript{-1}, not shown).
This is at variance with the simulation shown by~\cite{Arnold2012}, which exhibit an abrupt transition to superrotation for similar parameter values (their Fig. 6 shows a $k_f=1$ forcing with slow propagation $c_f=5$ m.s\textsuperscript{-1}, but they report similar behavior for stationary forcing over multiple forcing wave numbers), albeit with a different model, (CAM).
On the other hand, it is in agreement with the results of~\cite{Lutsko2018}, who found a reversible transition to superrotation, using the same dynamical core as the current study.
We note that the zonally-averaged zonal wind in the upper tropical troposphere is more sensitive to diabatic heating with larger zonal scale: the slope of the curve is largest for $k_f=1$ and decreases with $k_f$.
For $k_f=3$ and $k_f=4$ the response of the wind to the forcing is almost identical, and saturates to positive values around 5 m.s\textsuperscript{-1} corresponding to weak westerlies.
Extending the runs to larger forcing amplitudes, up to $Q_0=9$ K.day\textsuperscript{-1} (not shown), we find that the zonal-mean zonal wind starts increasing again, as reported by~\cite{Lutsko2018}, albeit less abruptly.
For $k_f=1$ the wind speed in the superrotating jet reaches values in excess of 30 m.s\textsuperscript{-1} and does not exhibit signs of saturation for $Q_0=2$ K.day\textsuperscript{-1} and above: it keeps increasing for forcing amplitudes up to $Q_0=9$ K.day\textsuperscript{-1} (not shown).

The absence of hysteresis in the above experiments is also in contradiction with the simple theoretical argument given by~\citet{Arnold2012} and~\citet{Herbert2020}  based on wave-jet resonance in a Matsuno-Gill framework.
A natural explanation could be that the resonant eddy acceleration induced by the non-zonal diabatic heating could be over-compensated by other acceleration terms resulting from aspects of the dynamics which are not accounted for in the Matsuno-Gill framework.
Hence we conduct similar hysteresis experiments in a setup which is closer to the Matsuno-Gill framework, by setting $\Delta_h=0$.
Simulations of this type were already conducted by~\citet{Arnold2012} with the CAM model, who found an abrupt transition in this case, and the fact that a larger meridional temperature gradient inhibits superrotation more generally was also pointed out by~\citet{Liu2010} and~\citet{Laraia2015}.
Our simulations with Isca also exhibit hysteresis in the absence of a meridional temperature gradient, as seen in Fig.~\ref{fig:hysteresis_dh} (right).
Unlike~\citet{Arnold2012}, we keep a meridionally varying lapse rate in the $\Delta_h=0$ case, so that there is a residual meridional temperature gradient; our results indicate that this does not affect the existence of hysteresis.
We note that in this case, the control run is already weakly superrotating, with tropical winds in the upper troposphere on the order of 1 m.s\textsuperscript{-1}.
When the diabatic heating increases, the wind decreases slightly (becoming weakly easterly again for the $k_f>1$ simulations) until a bifurcation is reached and the system undergoes an abrupt transition to a strongly superrotating state, characterized by upper-tropospheric tropical wind in excess of 25 m.s\textsuperscript{-1}.
The critical value of $Q_0$ for which the bifurcation occurs increases with the forcing wave number $k_f$, but it remains relatively small: below 0.25 K.day\textsuperscript{-1} for $k_f \leq 4$.
We have checked that larger forcing wave numbers behave similarly (not shown), with a slightly increasing bifurcation point (about 0.5 K.day\textsuperscript{-1} for $k_f=8$ for instance).
Although the response of zonal wind to non-zonal diabatic heating is clearly hysteretic, the decreasing $Q_0$ branch does not exhibit a clear bifurcation: the transition from the strongly superrotating state to a weakly superrotating state similar to the control run occurs smoothly as $Q_0$ is decreased.
A small discrepancy remains at $Q_0=0$ because these runs have not yet reached steady state.
The shape of the decreasing $Q_0$ branches is similar for different forcing wavenumbers, suggesting that the system follows a similar dynamics in some reduced space, regardless of the value of $Q_0$.

These results suggest that the wave-jet feedback mechanism is indeed present in our model, but that it does not suffice to generate hysteresis in the presence of a meridional temperature gradient.
We have conducted hysteresis experiments for several values of $\Delta_h$, which shows that for the default values of all other parameters, hysteresis is restricted to very low values of $\Delta_h$.
One can hypothesize that the positive feedback related to the Matsuno-Gill forcing is compensated by a stronger negative feedback in the presence of a meridional temperature gradient.
Theoretical study of the wave-jet feedback mechanism~\citep{Herbert2020} also suggests that the strength of the feedback, and ultimately the existence of hysteresis, is controlled by a parameter involving bottom friction, smaller friction leading to a stronger positive feedback.
In the next section, we test whether this result, obtained in the shallow-water setting, holds in our 3D model.

\subsection{Impact of bottom friction on the transition to superrotation}

\begin{figure*}[tbhp]\centering
  \includegraphics[width=0.48\linewidth]{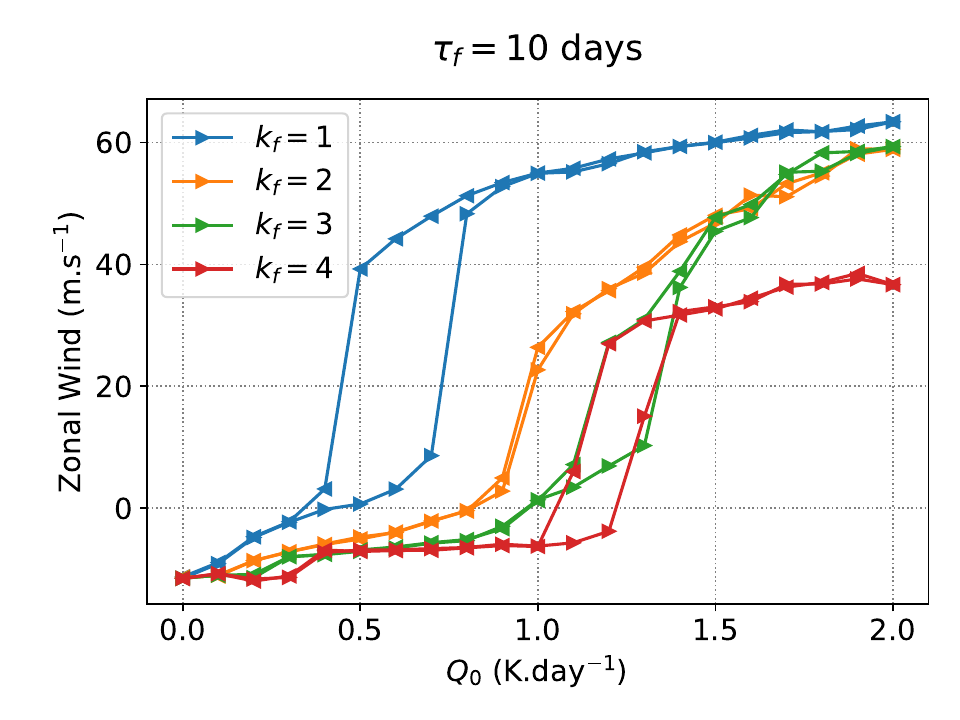}
  \includegraphics[width=0.48\linewidth]{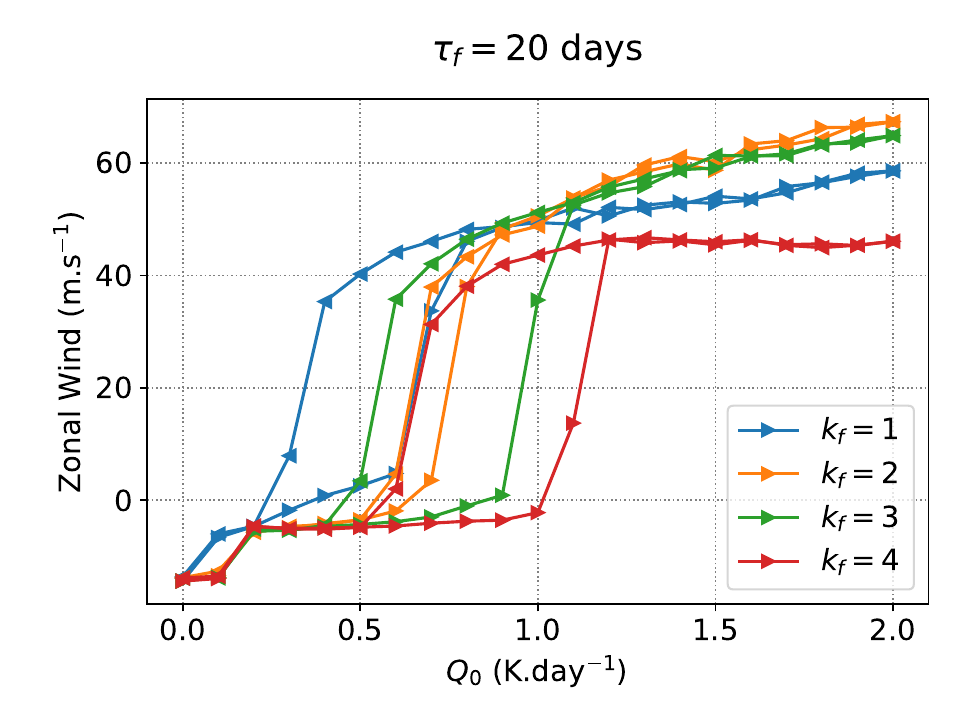}
  \caption{Zonally-averaged zonal wind in the upper troposphere at the equator as a function of the diabatic heating amplitude $Q_0$ with meridional temperature gradient $\Delta_h=40$ K and reduced bottom friction $\tau_f=10$ days (left) and $\tau_f=20$ days (right).}
  \label{fig:hysteresis_friction}
\end{figure*}

We carry out hysteresis simulations similar to the above section, with meridional temperature gradient $\Delta_h=40$ K, but with decreased friction: Fig.~\ref{fig:hysteresis_friction} shows the zonal-mean zonal wind at the equator as a function of the amplitude of the tropical diabatic heating $Q_0$ for $\tau_f=10$ days (left panel) and $\tau_f=20$ days (right panel), for multiple forcing wave numbers.
We conclude from these experiments that decreasing the friction coefficient indeed leads to an abrupt transition to superrotation, with clear hysteresis behavior for all the forcing wave numbers investigated.
We note that the value of the friction coefficient for which a bifurcation appears depends on the wave number: for $k_f=2$ there is no bifurcation at $\tau_f=10$ days, but it appears at $\tau_f=20$ days.
In addition, the value of $Q_0$ for which the bifurcation occurs depends on both the forcing wave number and the friction coefficient.
A general behavior is that the loss of stability of the subrotating or weakly superrotating state (the lower branch of the hysteresis diagram) occurs for smaller forcing amplitude at larger $\tau_f$.
The same holds for the bifurcation corresponding to loss of stability of the strongly superrotating state (the upper branch of the hysteresis diagram).
The bistability range, which is the difference between these two bifurcation points, also seems to increase with increasing $\tau_f$.

\begin{figure*}[tbhp]
  \centering
  \includegraphics[width=0.48\linewidth]{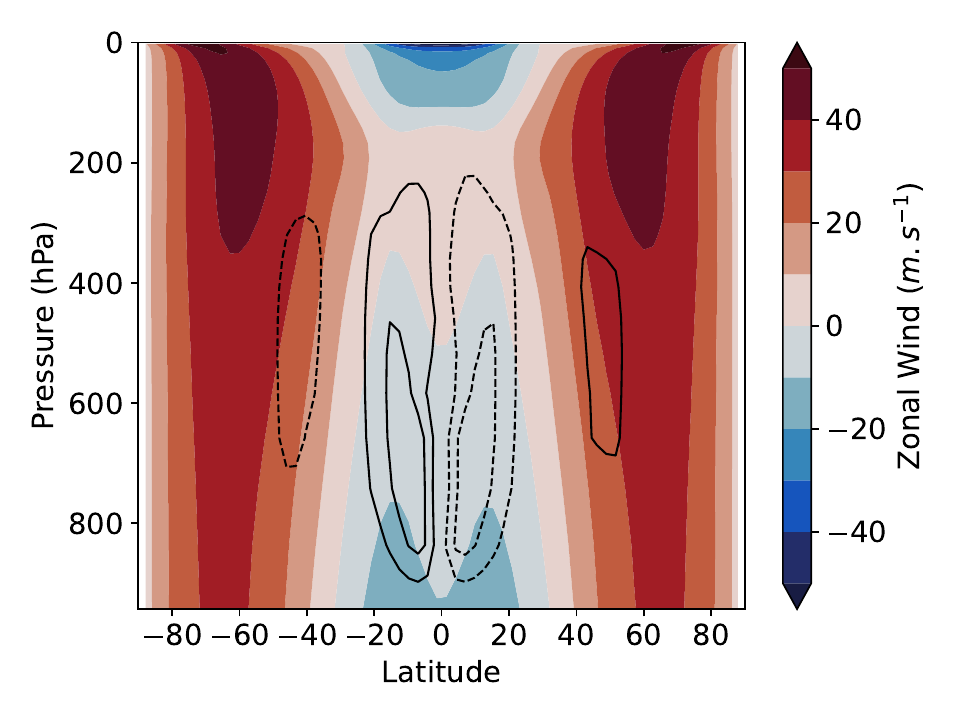}
  \includegraphics[width=0.48\linewidth]{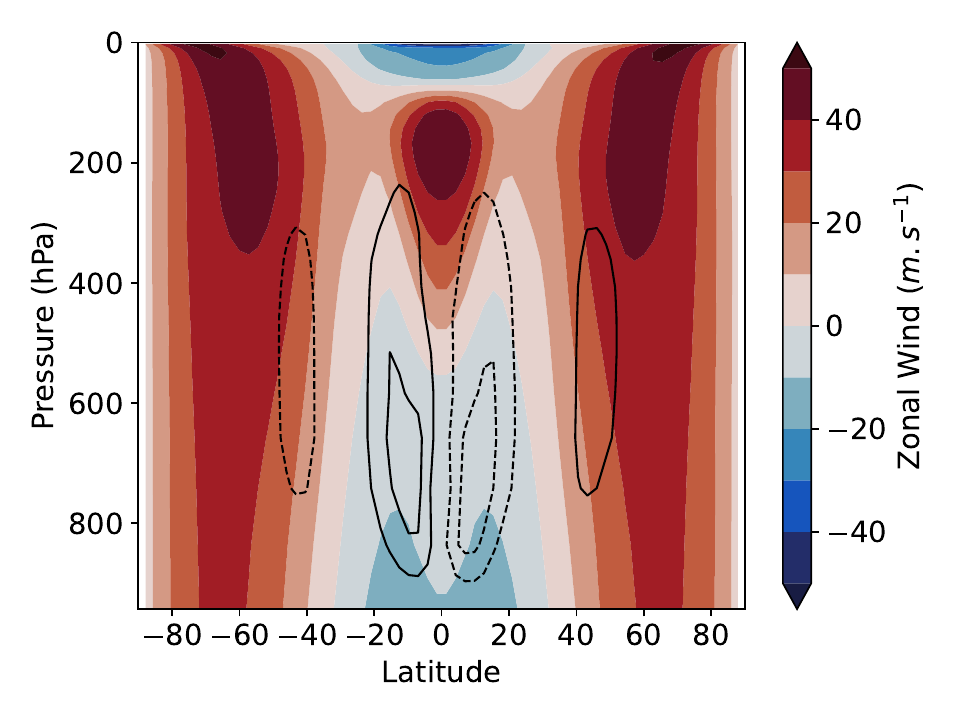}
  \caption{Zonally-averaged zonal wind (shading) and meridional mass streamfunction (contours) in the two coexisting states, sub-rotating (left) and superrotating (right), for the same value of the tropical diabatic heating amplitude $Q_0=0.6$ K.day\textsuperscript{-1}, with meridional temperature gradient $\Delta_h=40$ K and friction time $\tau_f=10$ days. The other forcing parameters are $k_f=1, c_f=0$. The contour interval for the meridional mass streamfunction is 10\textsuperscript{10} kg.s\textsuperscript{-1}.}
  \label{fig:circulation_friction}
\end{figure*}
We show in Fig.~\ref{fig:circulation_friction} the circulation associated with the two coexisting states, here in the $k_f=1$, $\tau_f=10$ days, $Q_0=0.6$ K.day\textsuperscript{-1} case.
Because of the low friction, the circulation, even in the absence of tropical heating, differs significantly from the control run: the jets have moved poleward and have a more barotropic structure~\citep{James1986,Robinson1997,Chen2007}.
Nevertheless, Fig.~\ref{fig:circulation_friction} shows that the two coexisting stable states have a markedly different tropical circulation: one exhibits weak westerlies in the upper tropical troposphere, while the other exhibits a strong eastward jet in this region.
Note that the mean meridional circulation is similar in both cases, with a weak Hadley cell below the region where the jet appears in the superrotating state.

Bottom friction is not the only dissipative mechanism here: the parameterization of radiative fluxes as a Newtonian relaxation introduces a radiative cooling mechanism which linearly damps the temperature field.
This is analogous to the shallow-water setup considered by~\citet{Herbert2020} for theoretical study of the wave-jet resonance mechanism, but in that study the time scales associated to these two mechanisms were assumed to be the same, while they can be controlled independently in the current simulations.
\citet{Wang2021a} have extended the work of~\citet{Herbert2020} to show that the wave-jet resonance mechanism still exists over a broad range of values when the time scales for friction and radiative cooling differ, in the shallow-water model.
Here we have checked that increasing the radiative cooling time scales $\tau_a$ and $\tau_s$ instead of the friction time $\tau_f$ also leads to hysteresis behavior.
In the next section, however, we restrict ourselves to studying the role of bottom friction on the feedback mechanisms governing the existence of a bifurcation.

\section{Feedback mechanisms controlling the existence of bifurcations}\label{sec:mechanisms}

\subsection{Angular momentum budget}\label{sec:ambudget}

To understand the role of different dynamical mechanisms in the transition to superrotation, we first consider the vertically and zonally-averaged angular momentum budget:
\begin{equation}
  \label{eq:am_budget}
  \frac{\partial \lbrack \bar{m}\rbrack}{\partial t} = - \frac{1}{a\cos\phi}\frac{\partial}{\partial\phi} \lbrack \bar{m}\bar{v}\cos\phi\rbrack - \frac{1}{a\cos\phi}\frac{\partial}{\partial \phi} \lbrack \overline{m'v'}\cos\phi\rbrack + \lbrack \bar{F}_u \rbrack a \cos\phi,
\end{equation}
where $m=(\Omega a\cos\phi+u)a\cos\phi$ is the total angular momentum, $a$ the Earth radius, $\bar{\cdot}$ the zonal average and $\lbrack\cdot\rbrack$ the mass-weighted vertical average.
In this equation, the angular momentum budget is affected by three effects: transport by the mean meridional circulation (first term on the right-hand side), eddy angular momentum fluxes (second term), and boundary layer friction (third term).
We can further decompose the eddy angular momentum fluxes into their stationary and transient components. 

\begin{figure*}[tbhp]
  \centering
  \includegraphics[width=\linewidth]{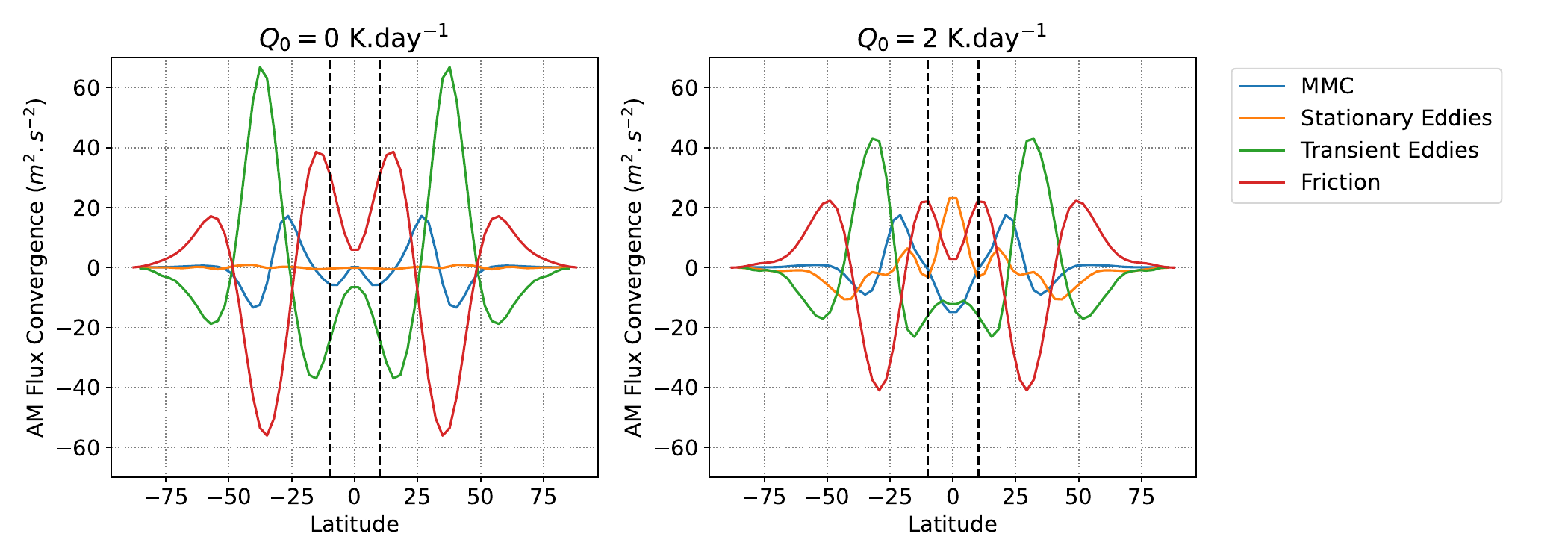}
  \caption{Zonally- and vertically-averaged angular momentum flux convergence as a function of latitude, in the control run ($\Delta_h=40$ K, $\tau_f=1$ day, $Q_0=0$  K.day\textsuperscript{-1}, left) and in a superrotating state (same parameters, $Q_0=2$ K.day\textsuperscript{-1}, $k_f=2$, $c_f=0$, right). The dashed vertical lines correspond to the latitudes 10S and 10N.} 
  \label{fig:am_profile}
\end{figure*}
The steady-state, zonally- and vertically-averaged angular momentum budget as a function of latitude is shown in Fig.~\ref{fig:am_profile} for the default values of all forcing parameters.
For the control run (left panel, $Q_0=0$), the main balance is between friction, which imparts angular momentum to the atmosphere in the tropics, and transient eddies, which transport the angular momentum to the mid-latitudes, where friction returns it to the planet.
The mean meridional circulation plays a subdominant role overall but is nevertheless responsible for angular momentum convergence at the poleward edge of the Hadley cell.
Stationary eddies are very weak in the control run and therefore do not play any role on the angular momentum budget, as could be expected from the fact that the boundary conditions are fully axisymmetric in this case.
In the superrotating state (right panel, $Q_0=2$ K.day\textsuperscript{-1}), the magnitude of the angular momentum flux convergence due to friction and transient eddies is reduced in the mid-latitudes and tropics, except very close to the equator, and the location of their maxima and minima, as well as those of the fluxes due to the mean meridional circulation are slightly shifted equatorwards.
The main difference with the control run is that the contribution of the stationary eddies is no longer negligible, and this affects in particular the main balance near the equator: now the dominant convergence term is the contribution of stationary eddies (friction still acts as an angular momentum source in this region but it is subdominant), balanced by the combination of transient eddies and mean meridional circulation, both of which act as stronger sinks than in the control run.

\begin{figure*}[tbhp]
  \centering
  \includegraphics[width=\linewidth]{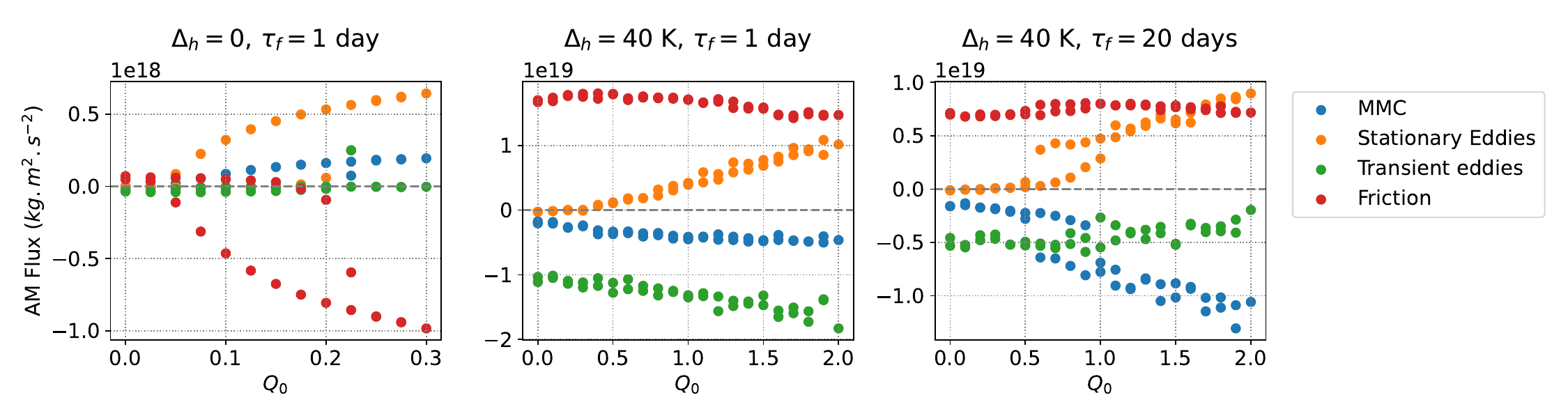}
  \caption{Zonally- and vertically-averaged angular momentum fluxes integrated over the 10S--10N latitude band (i.e. the region between dashed vertical lines in Fig.~\ref{fig:am_profile}), without meridional temperature gradient ($\Delta_h=0$, left), with meridional temperature gradient ($\Delta_h=40$ K) and standard friction ($\tau_f =1$ day, middle) or reduced friction ($\tau_f = 20$ days, right). The parameters for the tropical diabatic heating are $k_f=3$, $c_f=0$.}
  \label{fig:am_budget}
\end{figure*}

To describe more precisely the evolution of the angular momentum budget as the tropical diabatic heating increases, we show in Fig.~\ref{fig:am_budget} the same terms averaged over the latitude band 10S--10N, as a function of the amplitude of the tropical diabatic heating $Q_0$, with or without meridional temperature gradient and with standard or decreased bottom friction.
This confirms that for the default parameter values (middle panel) the main angular momentum budget changes when the diabatic heating amplitude $Q_0$ increases are that the stationary eddy angular momentum fluxes become an increasingly large source of angular momentum in the tropics.
This source is mostly compensated by an increased export of angular momentum by the transient eddies, and to a smaller extent by an increased export by the mean meridional circulation and a slightly decreased source due to bottom friction.
This analysis of the angular momentum budget does not allow one to conclude about potential positive or negative \emph{feedbacks} between the various angular momentum fluxes and the zonally-averaged zonal wind, because it does not take into account the changes of the zonal wind associated to the changes in diabatic heating amplitude $Q_0$.
However, it already points at the fact that as we tend to accelerate the mean zonal wind towards the east through the effect of stationary eddies by increasing the forcing, no other angular momentum source appears (which makes it plausible that the only potential positive feedback would be the stationary eddies), and the dominant term balancing this effect is the angular momentum sink due to transient eddies (as the forcing does not \emph{a priori} force directly these eddies, it suggests that the increased angular momentum flux out of the tropics is due to the interaction with the mean-flow, potentially through a negative feedback).
Before discussing in more details these two feedbacks in sections~\ref{sec:mechanisms}.\ref{sec:positivefeedback} and~\ref{sec:mechanisms}.\ref{sec:negativefeedback}, we consider the effect of the meridional temperature gradient and bottom friction on the angular momentum budget.

When the meridional temperature gradient is reduced to 0 (Fig.~\ref{fig:am_budget}, left panel), the tropical angular momentum budget is very different.
In the control run ($Q_0=0$) friction still imparts angular momentum to the atmosphere in the tropics, which is exported towards higher latitudes by transient eddies and the mean meridional circulation, but with much smaller fluxes than for $\Delta_h=40$ K.
Here too, as the tropical diabatic heating amplitude $Q_0$ increases the stationary eddies transport angular momentum upgradient, towards the equator.
However, the balancing mechanism is very different: because of the strongly reduced baroclinicity, the transient eddy angular momentum fluxes remain small at all forcing amplitudes.
Angular momentum transport by the mean meridional circulation not only remains significantly smaller than the stationary eddy fluxes: surprisingly it constitutes a net source of angular momentum in the tropics.
Hence, the only mechanism left to balance the budget is friction.
Unlike the standard picture for a sub-rotating atmosphere, in the superrotating state the eastward jet is deep enough to reach the boundary layer (which in this idealized setup corresponds to $\sigma=0.7$), and this is where the angular momentum is removed from the tropical atmosphere.

Going back to the control meridional temperature gradient $\Delta_h=40$ K, we now reduce the friction coefficient by increasing the friction time to 20 days.
As discussed in Sec.~\ref{sec:parameters}, this modifies the nature of the transition to superrotation and leads to bifurcations and hysteresis.
The angular momentum budget (Fig.~\ref{fig:am_budget}, right) is also affected.
Without tropical diabatic heating ($Q_0=0$), the budget is qualitatively the same as for the standard friction coefficient, only with smaller eddy fluxes, as expected for instance from~\cite{James1986}.
However, the behavior of these fluxes as the tropical diabatic heating amplitude $Q_0$ increases is very different.
Indeed, the export of angular momentum out of the tropics by transient eddies no longer increases with $Q_0$, and even decreases slightly for larger values of $Q_0$.
This suggests that if there is indeed a negative feedback mechanism between transient eddy angular momentum fluxes and zonally-averaged zonal wind, it is no longer active in the weak friction limit.
As a consequence, the mechanism which closes the budget is the transport by the mean meridional circulation, which becomes the dominant angular momentum sink in the tropics for large tropical diabatic heating.
We also note that the relative value of the angular momentum source due to stationary eddies increases compared to the standard friction case (it becomes the dominant source, larger than friction, at the highest values of $Q_0$), but this in itself does not say anything about a potential impact of friction on a putative positive-feedback mechanism.
In the next section, we discuss in more details this possibility.

\subsection{The positive wave-jet feedback}\label{sec:positivefeedback}

The notion of feedback can be formalized in an eddy-mean interaction framework, where we assume a timescale separation between the eddies and the mean.
When this condition is met, the properties of eddies (which can be either deterministic or stochastic) adjust to the mean-flow over a timescale much shorter than the timescale over which the mean-flow evolves, and we can think of the eddy terms in the mean-flow equation as unknown functions of the mean-flow.
Then, this function encodes the notion of feedback through its monotonicity: a positive feedback corresponds to an eddy term increasing when the mean-flow increases, and vice-versa for a negative feedback.
Here, we attempt to identify feedbacks by relying on this conceptual picture, and we shall plot various eddy momentum (or equivalently, angular momentum) flux terms as functions of the mean-flow.
A practical difficulty, however, is that such curves are not straightforward to build, because we do not control the mean-flow independently.
Let us take as an example the natural quantity to study the proposed positive feedback between stationary eddies forced by the tropical diabatic heating and the mean-flow.
According to quasi-linear theory~\citep{Herbert2020}, we expect the eddy momentum flux convergence to scale like the square of the heating amplitude $Q_0$.
Hence it is natural to plot $-\partial_y\overline{u'v'}/Q_0^2$ as a function of the mean flow $U=\bar{u}$ (averaged over the relevant region).
However, doing so with the stationary states of the hysteresis experiments described above yields only a few points on the curve we are trying to construct, and misses altogether the region corresponding to the potential positive feedback.
Hence, we will represent transient data where the mean-flow varies slowly, assuming that the eddy terms adjust fast enough so that the data points are in the vicinity of the idealized curve we expect to exist in the eddy momentum flux convergence/zonal-mean zonal wind plane.
The relevance of this picture can be assessed a posteriori by considering to what extent the data indeed seems to collapse onto a curve.

One problem raised in this approach is the question of the definition of stationary eddies, as it is no longer feasible simply to average in time in a steady-state.
A natural approach would be to do a running mean over some time window.
Here we suggest an alternative approach which yields relatively clean data.
\begin{figure*}[tbhp]
  \centering
  \includegraphics[width=\linewidth]{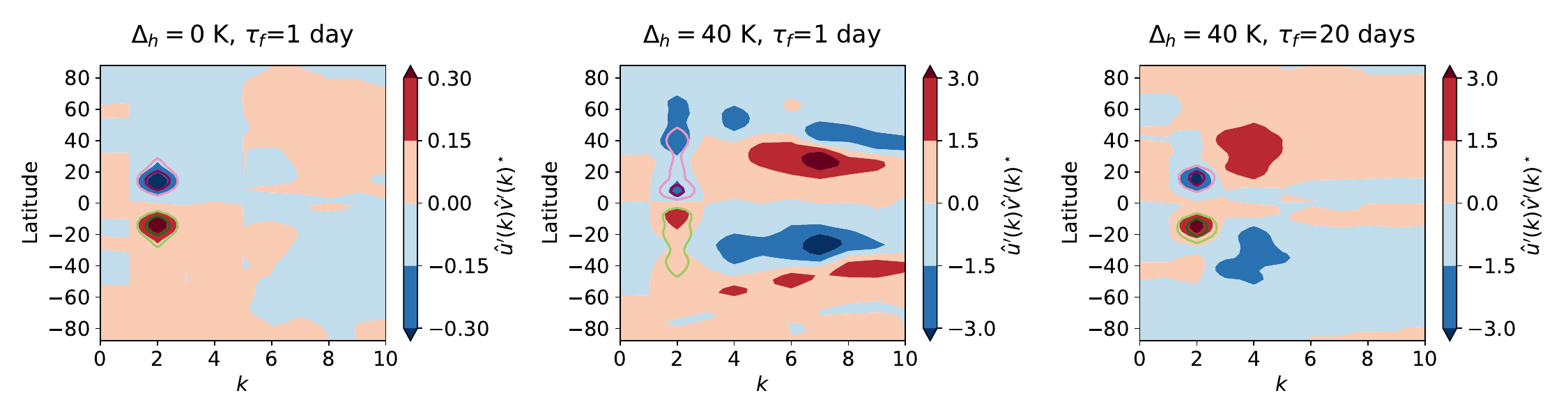}
  \caption{Latitude-zonal wave number spectrum of the eddy momentum flux (shading) and the same quantity for stationary eddies identified using a time-average (contours), for three cases: default parameters ($\Delta_h=40$ K, $\tau_f=1$ day, middle panel), reduced meridional temperature gradient ($\Delta_h=0$ K, left panel) and reduced friction coefficient ($\tau_f=20$ days, right panel). All the panels have forcing parameters $k_f=2$, $c_f=0$. With meridional temperature gradient $\Delta_h=40$ K (middle and right panels) the figures correspond to tropical diabatic heating amplitude $Q_0=1$ K.day\textsuperscript{-1} and for $\Delta_h=0$ (left panel), $Q_0=0.3$ K.day\textsuperscript{-1}. The cross-spectra are averaged over time and over the upper troposphere, between 100 hPa and 500 hPa.}\label{fig:emfc_spectrum}
\end{figure*}
We show in Fig.~\ref{fig:emfc_spectrum} the latitude-zonal wave number spectrum of the eddy momentum flux, $\hat{u}'(k)\hat{v}'(k)^\star$ where $\hat{\cdot}$ represents the Fourier transform in the longitude direction, so that the eddy momentum flux is given by $\overline{u'v'}=\sum_k \hat{u}'(k)\hat{v}'(k)^\star (\Delta{}k)^2$.
The value of the tropical diabatic heating amplitude is chosen so that the eddy fluxes are strong enough to be clearly visible.
As a consequence all the panels correspond to superrotating states.
With the default parameters ($\Delta_h=40$ K, $\tau_f=1$ day; center panel), we can clearly identify a range of eddies with wave numbers ranging from 4 to about 10, with a maximum around 7, transporting momentum polewards in the tropics.
These eddies are always present, regardless of the value of $Q_0$; their main source is the baroclinic instability of the extratropical jet, located at the latitude where they converge momentum.
In addition, we can also see eddies with wavenumber 2, which is also the wavenumber of the tropical diabatic heating, transporting momentum equatorwards.
These eddies are absent when $Q_0=0$, and their amplitude increase with $Q_0$: they correspond to the stationary eddies forced by the heating~\citep{Arnold2012,Herbert2020}, in agreement with the standard picture of~\citet{Matsuno1966} and~\citet{Gill1980}.
Indeed, we recover very similar features in the spectrum of the eddy momentum flux convergence in the case without meridional temperature gradient ($\Delta_h=0$, Fig.~\ref{fig:emfc_spectrum} left panel).
In that case, there is no other significant type of eddies as the baroclinic instability has been switched off.
We confirm that these features are mostly due to stationary eddies by plotting on top of the shading the contours of the eddy momentum flux convergence computed with time-averaged fields.
Finally, we can also consider the case with meridional temperature gradient but reduced friction ($\tau_f=20$ days, Fig.~\ref{fig:emfc_spectrum} right panel): it is similar to the default parameter values case, except that the eddies transporting westerly tropical momentum polewards have a smaller wave number and reduced amplitude, while the stationary eddies generated by the tropical diabatic heating have stronger amplitude.
This supports our hypothesis about the relative roles of these two mechanisms on the angular momentum budget when friction decreases.
To study the feedback mechanism in the way outlined above, we will rely on this spectral analysis to identify the stationary eddy momentum fluxes with spectrally filtered eddy momentum fluxes where we retain only wave numbers smaller than 4.
In other words, rather than the stationary eddy momentum flux $\overline{\langle{}u'\rangle\langle{}v'\rangle}$, where $\langle\cdot\rangle$ denotes the time average, we compute $(\overline{u'v'})_s=\sum_{k\leq 4} \hat{u}'(k)\hat{v}'(k)^\star (\Delta{}k)^2$.
These fluxes can be easily computed even in transient experiments.

\begin{figure}[tbhp]
  \centering
  \includegraphics[width=0.5\linewidth]{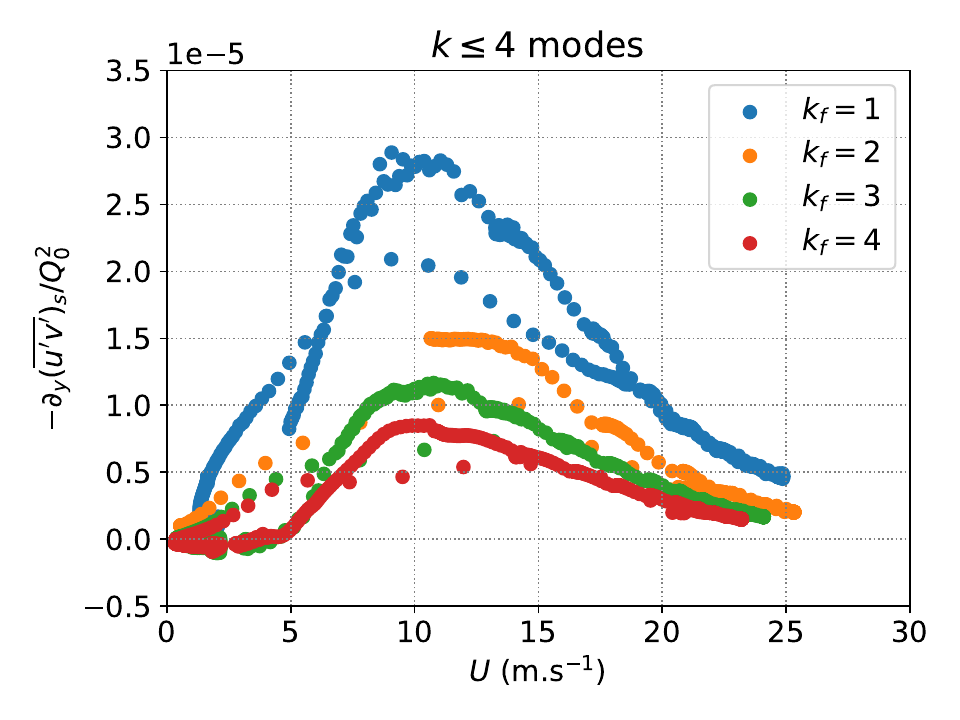}
  \caption{Normalized, spectrally-filtered eddy-momentum flux convergence, averaged over the upper tropical troposphere (between 5S and 5N and between 100 hPa and 500 hPa), as a function of zonal-mean zonal wind (averaged over the same region), for $\Delta_h=0$ and for forcing wave numbers $k_f=1$, $2$, $3$ and $4$.}
  \label{fig:resonance_dh00}
\end{figure}
We first try to identify the positive feedback due to forced tropical waves by showing in Fig.~\ref{fig:resonance_dh00} the so-defined stationary eddy momentum flux convergence as a function of the mean flow U for $\Delta_h=0$.
We note that the points in the figure do not correspond perfectly to a function of $U$: discrepancies are particularly important in the range of zonal-mean zonal wind values which do not correspond to stable steady states, as can be expected because close to the bifurcations the mean-flow evolves fast and the eddy terms are not in equilibrium with the instantaneous mean-flow.
Nevertheless, the general shape of the scatter plot indicates unequivocally a resonance curve, with a clear positive feedback for zonal-mean zonal winds below about 10 m.s\textsuperscript{-1}, and a negative feedback for larger velocities.
While this resonant behavior is consistent with the theoretical argument given by~\cite{Arnold2012} and~\cite{Herbert2020}, a number of differences appear.
Indeed, shallow-water theory predicts a resonance peak centered on the phase speed of Rossby waves, with a width on the order of $\alpha/k_f$ with $\alpha$ the friction coefficient.
In our primitive equations simulations, neither the position nor the width of the resonance peak seem to depend significantly on the wave number.
In addition, the curves obtained for decreased friction (e.g. $\tau_f=20$ days, not shown) are virtually indistinguishable from the ones shown in Fig.~\ref{fig:resonance_dh00}, again in contradiction with shallow-water theory (according to which the resonance peaks should be narrower).
One could hypothesize that the width of the resonance is controlled by another dissipative mechanism, such as radiative cooling.
However, even though bottom friction does not act in the region where the diabatic heating is maximum, the vertical structure of the stationary eddies forced by this heating is deep enough to reach the boundary layer, and friction does have an impact on the spatial structure of the eddies.
In fact, both friction and radiative cooling independently modulate the amplitude of the stationary eddies for a given diabatic heating amplitude: the eddies are slightly stronger for weaker friction, and slightly weaker for weaker radiative cooling.
The tilt and vertical profile of the eddies is also affected in both cases.
\citet{Wang2021a} have generalized the shallow-water theory to distinguish the effect of friction and radiative cooling.
While they find that the peak in the eddy-momentum flux convergence at the equator needs not coincide with the Rossby wave phase speed in general, their results indicate that it tends to move towards larger values of the zonal-mean zonal wind and become narrower when friction decreases.
This behavior is not observed in our 3D primitive equations simulations.

\begin{figure}[tbhp]
  \centering
  \includegraphics[width=0.5\linewidth]{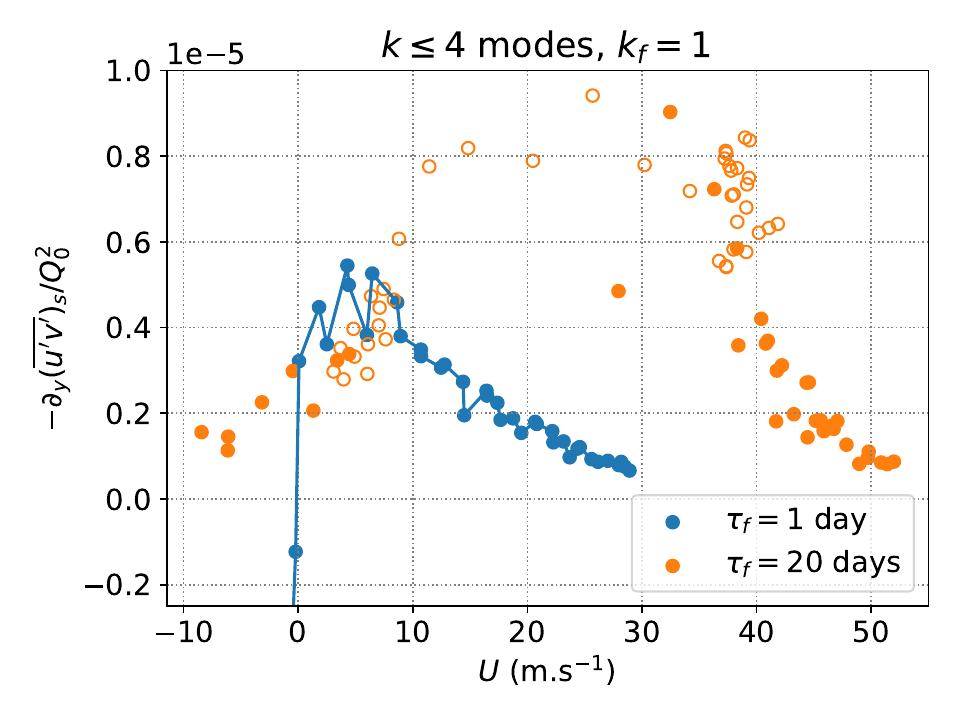}
  \caption{Same as Fig.~\ref{fig:resonance_dh00} but for $\Delta_h=40$ K. We only show a single forcing wave number ($k_f=1$) but for two values of the friction time: $\tau_f=1$ day and $\tau_f=20$ days. Filled dots represent data averaged over the full duration of the segment of the hysteresis experiment (5000 days), while open dots are time-averaged with a 450-day sliding window.}
  \label{fig:resonance_dh40}
\end{figure}
We plot in Fig.~\ref{fig:resonance_dh40} the same spectrally-filtered stationary eddy momentum flux convergence as a function of zonal-mean zonal wind, averaged over the same tropical region, for $\Delta_h=40$ K.
In the reference friction case ($\tau_f=1$ day), the shape of the curve is quite different from the $\Delta_h=0$ case.
Indeed, in the relevant range of zonal-mean zonal wind values (positive values), this term essentially always behave as a negative feedback, i.e. the curve decreases with $U$.
This is compatible with the reasoning that the lack of bistability in this case is at least partly due to a weak or nonexistent positive wave-jet feedback.

When friction is decreased (which restores bistability), it is more difficult to plot a resonance curve like above, because all the terms in the acceleration budget fluctuate much more.
Hence, we apply some time-averaging to smooth the data.
Away from the bifurcation, it is sufficient to average over the whole duration of the simulation leg (filled dots).
On the other hand, to sample the region separating the steady states, we need to retain some transient information.
Hence, for the segment of the hysteresis experiment with forcing amplitudes just above or below the bifurcation, we apply some time averaging with a sliding window (open dots), the length of which is empirically chosen as 450 days.
Combining the two types of data points reveals a curve with the shape of a resonance peak again, consistent with the interpretation that decreasing friction increases the positive wave-jet feedback.
Like in the $\Delta_h=0$ K case, the difference between the weak and strong friction cases for $\Delta_h=40$ K is only partially explained by shallow-water theory.
Indeed, while \citet{Arnold2012} and~\citet{Herbert2020} argue based on quasi-linear computations with a single-layer divergent model that the role of friction should be to make the resonance narrower, which~\citet{Wang2021a} later confirmed to be expected in such a model even when friction and radiative cooling differ, this is not clearly observed in the GCM simulations presented here.
On the other hand, the main effect of friction clearly visible in Fig.~\ref{fig:resonance_dh40} is to shift the position of the resonance peak, as the maximum of the stationary eddy momentum flux convergence moves to much larger zonal-mean zonal wind values when friction is decreased to $\tau_f=20$ days.
This is in accordance with the shallow-water results of~\citet{Wang2021a}.
Here, it seems to be a crucial factor to allow for the existence of bistability, as it establishes a relatively broad range of background velocities over which a positive wave-jet feedback operates.

\subsection{The negative feedback: extra-tropical wave absorption near critical lines}\label{sec:negativefeedback}

The analysis of the angular momentum budget (Sec.~\ref{sec:mechanisms}.\ref{sec:ambudget}) and its evolution with the tropical diabatic heating amplitude has suggested that transient eddies might play the role of a negative feedback which prevents an abrupt transition to superrotation for the standard parameter values (in particular, bottom friction).
A natural hypothesis is that the absorption of extratropical waves might act as a friction on the mean-flow in the tropics.
This hypothesis is supported by the fact that removing the source of extratropical waves by setting $\Delta_h=0$ leads to the existence of a bifurcation between the sub- and super-rotating states (Sec.~\ref{sec:parameters}.\ref{sec:meridional_gradient}).
However, it remains to better understand the connection between the strength of this friction and the strength of the mean-flow at the equator, i.e. the feedback, and why this feedback might be reduced or removed altogether when bottom friction is decreased.
In this section, we focus on the $\Delta_h=40$ K case.

\begin{figure*}[tbhp]
  \centering
  \includegraphics[width=\linewidth]{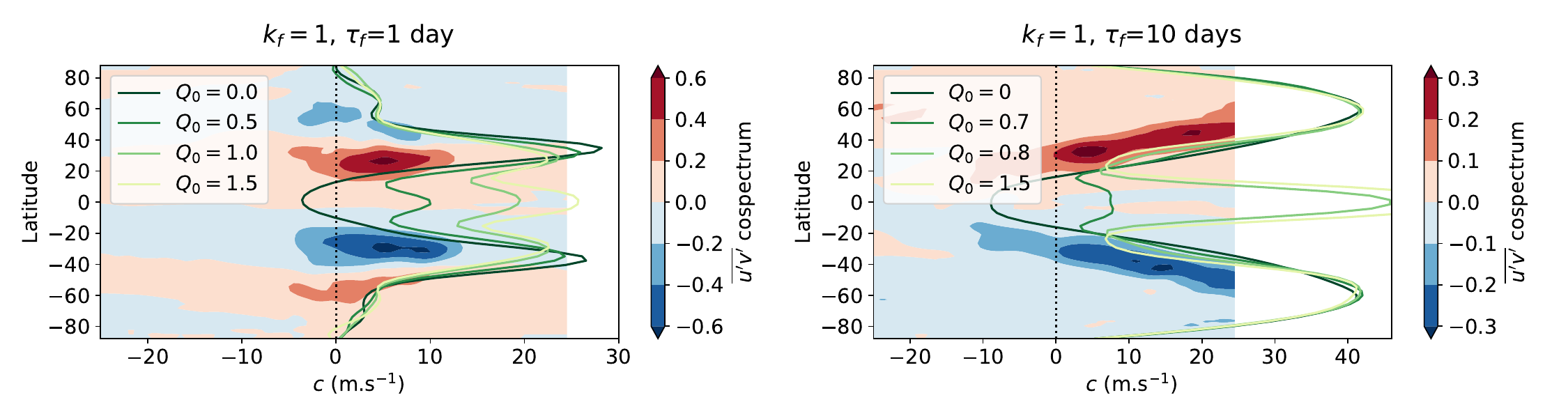}
  \caption{Latitude-phase speed spectra for the eddy momentum flux (shading) at 300 hPa in the simulations with $\Delta_h=40$ K, $k_f=1$, for two different friction times: the reference value $\tau_f=1$ day (left, for $Q_0=0.5$ K.day\textsuperscript{-1}) and $\tau_f=10$ days (right, for $Q_0=0.7$ K.day\textsuperscript{-1}). The green lines show the zonal-mean zonal wind profile at 300 hPa with increasingly light shades for increasing values of $Q_0$.}
  \label{fig:phasespeed_spectra}
\end{figure*}
To do so, we compute the latitude-phase speed spectrum of the eddy momentum flux $\overline{u'v'}$ following the standard method of~\citet{Randel1991}.
In Fig.~\ref{fig:phasespeed_spectra}, we show this spectrum in the upper troposphere (300 hPa) for both the reference bottom friction parameter ($\tau_f=1$ day), for which the transition to superrotation is smooth, and decreased bottom friction ($\tau_f=10$ days), for which a bifurcation occurs.
Note that the figure shows the spectrum only for a single value of the tropical diabatic heating amplitude in each case: indeed, the main features of this latitude-phase speed spectrum are independent of $Q_0$, and the small changes due to the tropical diabatic heating do not affect the discussion here.
The main effect of the heating occurs in the $\tau_f=10$ days case, where it displaces slightly poleward the regions of poleward eddy momentum fluxes on the equatorward flanks of the eddy-driven jets.
For $\tau_f=10$ days, we choose a value of $Q_0$ close to the bifurcation where the sub-rotating state looses stability, $Q_0=0.7$ K.day\textsuperscript{-1}, and for the $\tau_f=1$ day, we choose a similar value, $Q_0=0.5$ K.day\textsuperscript{-1}.
On top of these spectra, the figure shows the meridional profile of the zonal-mean zonal wind at the same level, 300 hPa, for increasing values of the tropical diabatic heating amplitude $Q_0$.

This diagnostic reveals a major difference between the reference friction case ($\tau_f=1$ day) and the decreased friction ($\tau_f=10$ days).
In the control run ($Q_0=0$) for each case, we recover the standard picture with eddy momentum flux convergence near the latitude of maximum zonal-mean zonal-wind (in fact, the maximum convergence is located slightly poleward of the wind maximum for the reference friction, and slightly equatorward for decreased friction) and divergence on the equatorward flank of the jet.
In particular, for each phase speed, the eddy momentum flux spectrum is non-zero only between critical lines for which the zonal-mean zonal wind speed matches the phase speed, suggesting that extratropical waves cannot propagate past these critical lines and are absorbed at or near these critical lines where they deposit their momentum.
When the mean-flow changes under the action of the tropical stationary eddies forced by the diabatic heating, the interplay with the extratropical waves differs for the two cases.

For the reference friction case, the tropical wind profile accelerates more or less uniformly when the diabatic heating amplitude increases: a local maximum is created at the equator, bu the two local minima at about 10S and 10N also move with the equatorial maximum.
Because of this coupling, as the tropical zonal wind becomes more westerly, it undergoes an easterly acceleration due to extratropical wave absorption near the critical lines of the eddies with phase speed in the range of the tropical wind profile.
The tropics become progressively transparent to a broader range of waves, but there always remain some critical lines damping the tropical wind.
As the diabatic heating amplitude increases and the tropical zonal wind profile enters the range of phase speeds with larger eddy momentum flux (between about 0 and 10 m.s\textsuperscript{-1}), this damping effect becomes larger.
With decreased bottom friction, a similar behavior is expected only in the first stage, up to a certain value of the diabatic heating amplitude (around 0.7 K.day\textsuperscript{-1} in the case shown on the figure).
Beyond this point, the local minimum in the zonal-mean zonal wind profile becomes independent of the wind at the equator, and the maximum grows without much changes in the off-equatorial minima (if anything, they move slightly poleward).
In this regime, the equatorial wind is effectively shielded from the extratropics: extratropical waves continue to deposit easterly momentum near their critical lines, but these critical lines do not move jointly with the equatorial wind.
This scenario suggests an interpretation for the disappearance of the negative feedback due to transient eddies in the low friction limit.

\section{Conclusion}\label{sec:conc}

The goal of this work was to investigate the nature of the transition to superrotation, smooth or abrupt, in a dry \ac{gcm} in an idealized Held-Suarez setup, allowing for a systematic study of the parameters controlling the behavior of the model.
Few previous studies have considered this question: also in a Held-Suarez setup,~\citet{Arnold2012} reported abrupt transitions to superrotation with hysteresis behavior in CAM with an imposed tropical diabatic heating, but this behavior was not found by~\citet{Lutsko2018} in a similar setup with the GFDL dynamical core.
Using the same model as~\citet{Lutsko2018}, we find that abrupt transitions with hysteresis are possible, but only in a region of the parameter space spanned by the meridional temperature gradient and the bottom friction coefficient.
This results from a competition between two mechanisms: a resonant tropical wave-jet mechanism, already studied by~\citet{Arnold2012} and~\citet{Herbert2020}, acting as a positive feedback, and tropical absorption of extratropical waves, acting as a negative feedback.
When the meridional temperature gradient is weak, the negative feedback is weak and abrupt transitions to superrotation with hysteresis are obtained.
This behavior is very robust and does not depend on other parameters such as bottom friction.
When the meridional temperature gradient increases, the negative feedback quickly becomes strong enough to overcome the positive feedback, resulting in a smooth, reversible transition to superrotation.
However, decreasing friction restores the hysteretic behavior.
The effect of reduced friction is twofold: it leads to a more efficient positive feedback, although not exactly in the way predicted by the shallow-water theory~\citep{Arnold2012,Herbert2020}, and simultaneously prevents the negative feedback by decoupling the subtropical minimum of the zonal wind profile from the equatorial jet, which shields the equatorial jet from the extratropical waves.
This behavior suggests that the properties of the control state, and in particular the mid-latitude wave spectrum and the related zonal wind profile, are key to determine whether the transition to superrotation is abrupt or continuous.

Given the degree of idealization of the setup, it is striking that two different models (CAM and GFDL) exhibit such a radically different transition to superrotation when subjected to a well-controlled forcing.
A natural follow-up to the present work would be to investigate in more details the difference between the response of various dynamical cores to such forcing in the light of the physical mechanisms uncovered here.
Another crucial point would be to relate the well-defined control parameters in the Held-Suarez setup to the parameterizations of fully-fledged \acp{gcm}, and to investigate whether the feedback processes identified here can be observed in such models as well.

We also note that the bifurcation structure of the system is more complex than what we have described here.
While we have clearly observed lines of saddle-node bifurcations in the $(\Delta_h, Q_0)$ plane, our simulations suggest the existence of other bifurcations. For instance, self-sustained oscillations of the upper-tropospheric zonal wind, reminiscent of the \ac{qbo}, are observed for $\Delta_h=10$ K, for the reference value of all the other parameters.
Another natural question is whether spontaneous transitions between sub-rotating and superrotating states could be observed in the parameter range corresponding to bistability.
In the simulations presented here, which are deterministic, this is not the case, at least not over 2000-year long stationary simulations starting from either of the two states.
One could expect to observe such transitions by adding a small-noise, but the statistical properties of the transitions have not been studied.
We leave the question of the detailed study of the bifurcation structure and spontaneous transitions for future studies.

%

%

\clearpage
\acknowledgments

The numerical simulations were carried out using the resources of the \emph{Pole Scientifique de Modélisation Numérique} \& \emph{Centre Blaise Pascal} at ENS de Lyon, and in particular the SIDUS system~\citep{Quemener2013}.
We are grateful to Cerasela Calugaru and Emmanuel Quemener for their help with the platform.
This work was supported by the ANR Grant \emph{TippingWinds}, Project No. ANR-21-CE30-0016-01, and by the \textit{Fondation Simone et Cino Del Duca - Institut de France}.

%
%
\datastatement

The simulations described in this paper were performed using the Isca model. The code, including the modifications necessary to implement the forcing used in the experiments reported here, is available at \url{https://github.com/cbherbert/Isca/tree/super}.

%






%



\bibliographystyle{ametsocV6}
\bibliography{bibliography}

\end{document}